\documentclass{appolb}
\usepackage{epsfig}

\begin{document}
\title{Thermodynamics of dense matter in chiral approaches%
\thanks{Presented at Excited QCD, Jan 31-Feb 6 2010,
Tatra National Park, Slovakia.}%
}
\author{Chihiro Sasaki
\address{Frankfurt Institute for Advanced Studies,
 D-60438 Frankfurt am Main,
 Germany}
}
\maketitle
\begin{abstract}
We discuss phases in dense hadronic and quark matter from chiral model
approaches. Within PNJL models the phase diagram for various number of
colors $N_c$ is studied. How phases are constrained in quantum field 
theories are also discussed along with the anomaly matching.
An exotic phase with unbroken center symmetry of chiral group has a 
characteristic feature in the thermodynamics, which can be interpreted 
as one realization of the quarkyonic phase in QCD for $N_c=3$.
\end{abstract}

  
\section{Phase diagram: from $N_c = \infty$ to $N_c=3$}
\label{sec:int}

Model studies of dense baryonic and quark matter have suggested 
a rich phase structure of QCD at temperatures and quark chemical 
potentials being of order $\Lambda_{\rm QCD}$.
Our knowledge on the phase structure is however still limited
and the description of the matter around the phase transitions 
does not reach a consensus because of the non-perturbative nature
of QCD~\cite{qmproc}.
Possible phases and spectra of excitations are guided by
symmetries and their breaking pattern in a medium.
Dynamical chiral symmetry breaking and confinement
are characterized by strict order parameters
associated with global symmetries of the QCD Lagrangian
in two limiting situations:
the quark bilinear $\langle \bar{q}q \rangle$ in the limit
of massless quarks, and the Polyakov loop $\langle \Phi \rangle$ 
in the limit of infinitely heavy quarks.

A novel phase of dense quarks, Quarkyonic Phase,
was recently proposed based on the argument using 
large $N_c$ counting where $N_c$ denotes number of 
colors~\cite{quarkyonic}:
in the large $N_c$ limit there are three phases which are
rigorously distinguished using $\langle \Phi \rangle$ and the 
baryon number density $\langle N_B \rangle$. The quarkyonic phase 
is characterized by $\langle \Phi \rangle = 0$ indicating the 
system confined and non-vanishing $\langle N_B \rangle$ above 
$\mu_B = M_B$ with a baryon mass $M_B$. 

\begin{figure}
\begin{center}
\includegraphics[width=8cm]{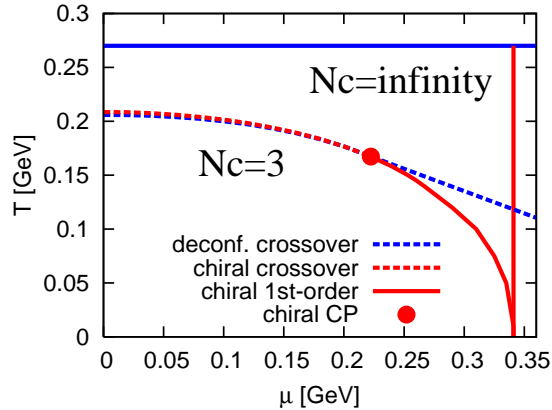}
\caption{
The phase diagram of a PNJL model for different $N_c$~\cite{mrs}. 
}
\label{pnjl}
\end{center}
\end{figure}

A possible deformation of the phase boundaries in large $N_c$
together with the chiral phase transition can be described
using a chiral model coupled to the Polyakov loop~\cite{mrs}. 
The Nambu--Jona-Lasinio model with Polyakov loops (PNJL model) 
has been developed to deal with chiral dynamics and ``confinement''
simultaneously~\cite{pnjl}.
The model describes that only three-quark states are thermally
relevant below the chiral critical temperature, which is
reminiscent of confinement.
Figure~\ref{pnjl} shows the two transition lines for $N_c=\infty$
and for $N_c=3$ in the two-flavored PNJL model.
In the large $N_c$ limit assuming that the system is
confined, the gap equations for the order parameters 
$\langle \bar{q}q \rangle$ and $\langle \Phi \rangle$ 
become two uncorrelated equations. Consequently,
the quark dynamics carries only a $\mu$ dependence
and the Polyakov loop sector does only a $T$ dependence.
Finite $N_c$ corrections make the transition lines bending down.
The crossover for deconfinement shows a weak dependence on
$\mu$ which is a remnant of the phase structure in large $N_c$.
One finds that for $N_c=3$ deconfinement and chiral crossover lines 
are on top of each other in a wide range of $\mu$. A critical point 
associated with chiral symmetry appears around the junction of those 
crossovers.

The clear separation of the quarkyonic from hadronic phase
is lost in a system with finite $N_c$.
Nevertheless, an abrupt change in the baryon number
density would be interpreted as the quarkyonic transition which 
separates meson dominant from baryon dominant regions~\cite{triple}.
In fact, a steep increase in the baryon number density and the 
corresponding maximum in its susceptibility $\chi_B$ are driven 
by a phase transition from chirally broken to restored phase 
in most model-approaches using constituent quarks.
One might then consider the chirally symmetric confined phase
as the quarkyonic phase.

The constituent quarks are however unphysical in confined phase.
It is not obvious to have a realistic description of
hadrons from chiral quarks.
In particular, chiral symmetry restoration for baryons must
be worked out. Two alternatives for chirality assignment are 
known~\cite{pdoubling} and it remains an open question which 
scenario is preferred by nature:
(i) in the naive assignment, dynamical chiral symmetry breaking
generates a baryon mass which thus vanishes at the restoration.
(ii) in the mirror assignment, dynamical chiral symmetry breaking
generates a mass difference between parity partners and the chiral
symmetry restoration does not necessarily dictate the chiral
partners being massless. If the chiral invariant mass is not
very small, the baryon number density is supposed to be insensitive
to the quarkyonic transition.

Besides, it seems unlikely that the chirally-restored confined phase
is realized in QCD on the basis of the anomaly matching:
external gauge fields, e.g. photons, interacting with quarks
lead to anomalies in the axial current. Since there are no
Nambu-Goldstone bosons in chiral restored phase, the anomalous
contribution must be generated from the triangle diagram
in which the baryons are circulating. In three flavors, however,
the baryons forming an octet do not contribute to the pole in the 
axial current because of the cancellations~\cite{shifman:anomaly}.
The mirror scenario has nothing to do with this problem because
the sign of the axial couplings to the positive and negative parity 
states are relatively opposite. It is indispensable to any rigorous
argument for this taking account of the physics around the Fermi 
surface, which could lead to a possibility of the chirally restored 
phase with confinement. The anomaly matching conditions at finite 
temperature and density are in fact altered, 
see e.g.~\cite{anomalymatching}.

\section{Role of the tetra-quark at finite density}
\label{sec:znf}

There is a possibility of two different phases with broken
chiral symmetry distinguished by the baryon number density.
An alternative pattern of spontaneous chiral symmetry breaking 
was suggested in the context of QCD at zero temperature and 
density~\cite{stern,SDE,SDE2}.
This pattern keeps the center of chiral group unbroken, i.e.
\begin{equation}
SU(N_f)_L \times SU(N_f)_R \to SU(N_f)_V \times (Z_{N_f})_A\,,
\label{breaking}
\end{equation}
where a discrete symmetry $(Z_{N_f})_A$ is the maximal axial
subgroup of $SU(N_f)_L \times SU(N_f)_R$. The center $Z_{N_f}$ 
symmetry protects a theory from condensate of quark bilinears 
$\langle \bar{q}q \rangle$. Spontaneous symmetry breaking
is driven by quartic condensates which are invariant under
both $SU(N_f)_V$ and $Z_{N_f}$ transformation.
Although meson phenomenology with this breaking pattern
seems to explain the reality reasonably~\cite{stern},
this possibility is strictly ruled out in QCD both at zero
and finite temperatures but at zero density since a different
way of coupling of Nambu-Goldstone bosons to pseudo-scalar
density violates QCD inequalities for density-density
correlators~\cite{shifman}.
However, this does not exclude the unorthodox pattern
in the presence of dense matter. 
In a system with the breaking pattern~(\ref{breaking})
the quartic condensate is the strict order parameter which
separates different chirally-broken phases~\footnote{
 A similar phase structure was discussed in~\cite{watanabe,skyrmion}.
}.



\begin{figure}
\begin{center}
\includegraphics[width=8cm]{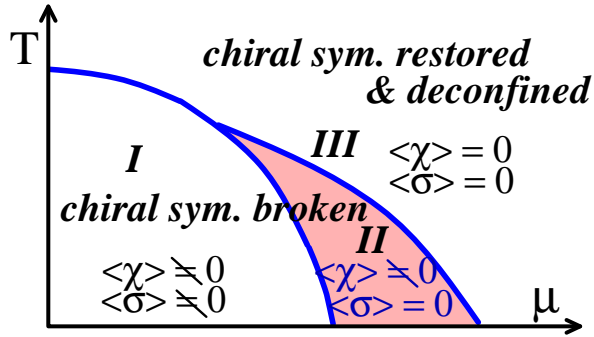}
\caption{
Schematic phase diagram mapped onto $(T,\mu)$ plane.
}
\label{phase}
\end{center}
\end{figure}

Assuming~(\ref{breaking})  
at finite density,
it has been shown that an intermediate phase between
chiral symmetry broken and its restored phases can be realized
using a general Ginzburg-Landau free energy~\cite{hst}.
The pion decay constant is read from the Noether current as
\begin{equation}
F_\pi = \sqrt{\sigma_0^2 + \frac{8}{3}\chi_0^2}\,,
\end{equation}
with $\chi_0$ and $\sigma_0$ being the expectation values
of 4-quark and 2-quark scalar fields, determined from the gap equations.
%
%
The effective potential deduced in the mean field approximation
describes three distinct phases characterized by the two order
parameters: Phase I represents the system where both chiral 
symmetry and its center are spontaneously broken due to 
non-vanishing expectation values $\chi_0$ and $\sigma_0$. 
The center symmetry is restored when $\sigma_0$ becomes zero. 
However, chiral symmetry remains broken as long as $\chi_0$ is non-vanishing,
where the pure 4-quark state is the massless
Nambu-Goldstone boson (phase II). The chiral symmetry restoration takes
place under $\chi_0 \to 0$ which corresponds to phase III.
The phases II and III are separated by a second-order line,
while the broken phase I from II or from III is by both first-
and second-order lines. 
Accordingly, 
there exist two tricritical points
(TCPs) and one triple point. One of these TCP
is associated with the center $Z_2$ symmetry
restoration rather than the chiral transition.
%
With an explicit breaking of chiral symmetry
one would draw a phase diagram mapped onto $(T,\mu)$ plane
as in Fig.~\ref{phase}.

Appearance of the above intermediate phase seems to have
a similarity to the notion of Quarkyonic Phase. 
The transition from hadronic to quarkyonic world
can be characterized by a rapid change in the net baryon number
density. In our model this feature is driven by the restoration 
of center symmetry and is due to the fact that the Yukawa coupling 
of $\chi$ to baryons is not allowed by the $Z_2$ invariance.
Consequently, the baryon number susceptibility exhibits a maximum
when across the $Z_2$ cross over.
This can be interpreted as the realization  
of the quarkyonic transition in $N_c=3$ world.
The phase with $\chi_0\neq0$ and $\sigma_0=0$ does not seem to
appear in the large $N_c$ limit~\cite{SDE2,shifman,watanabe}.
It would be expected that including $1/N_c$ corrections induce 
a phase with unbroken center symmetry.

\section{Conclusions}
\label{sec:conclusions}

We have discussed the phases in dense QCD from chiral approaches
along with the anomaly matching which is a field-theoretical
requirement. Although the chiral restored phase below ``deconfinement''
seems to be a common feature in PNJL models, this might be an artifact
of this toy model in which the temporal gluon field is treated as
a constant background and thus confinement dynamics is lost.
A possibility of a non-standard breaking pattern leads 
to a new phase where chiral symmetry is spontaneously broken 
while its center symmetry is restored. This might appear as an 
intermediate phase between chirally broken and restored phases 
in $(T,\mu)$ plane. The appearance of this phase also suggests 
a new critical point in low temperatures.
A tendency of the center symmetry restoration is carried by
the net baryon number density which shows a rapid increase
indicating baryons more activated,
and this is reminiscent of the quarkyonic transition.

The properties of baryons near the chiral phase transition
are also an issue to be clarified. Depending on the chirality
assignment to baryons, equations of state may be altered.
The chirality assignment becomes more involved when one
introduces axial-vector mesons~\cite{frankfurt}.
In this case the axial-vector meson gives a non-trivial contribution
to the axial couplings and eventually it is not clear that the sign
of the axial coupling to the negative parity state does distinguish
two scenarios.
In this respect, it attracts an interest
that the same sign of the axial couplings to the parity partners
is predicted in a top-down holographic QCD model~\cite{hss}
and the lattice QCD with dynamical quarks~\cite{takahashi}.

\section*{Acknowledgments}

I am grateful for fruitful collaboration with 
M.~Harada, L.~McLerran, K.~Redlich and S.~Takemoto.
Parital support by the Hessian
LOEWE initiative through the Helmholtz International
Center for FAIR (HIC for FAIR) is acknowledged.



\begin{thebibliography}{00}
\bibitem{qmproc}
C.~Sasaki,
  Nucl.\ Phys.\  A {\bf 830}, 649C (2009).

\bibitem{quarkyonic}
  L.~McLerran and R.~D.~Pisarski,
  Nucl.\ Phys.\  A {\bf 796}, 83 (2007),
  Y.~Hidaka, L.~D.~McLerran and R.~D.~Pisarski,
  Nucl.\ Phys.\  A {\bf 808}, 117 (2008).

\bibitem{mrs}
  L.~McLerran, K.~Redlich and C.~Sasaki,
  Nucl.\ Phys.\  A {\bf 824}, 86 (2009).

\bibitem{pnjl}
K.~Fukushima,
  Phys.\ Lett.\  B {\bf 591}, 277 (2004),
C.~Ratti, M.~A.~Thaler and W.~Weise,
  Phys.\ Rev.\  D {\bf 73}, 014019 (2006).

\bibitem{triple}
  A.~Andronic {\it et al.},
  Nucl.\ Phys.\  A {\bf 837}, 65 (2010).

\bibitem{pdoubling}
  C.~E.~Detar and T.~Kunihiro,
  Phys.\ Rev.\  D {\bf 39}, 2805 (1989),
  D.~Jido, M.~Oka and A.~Hosaka,
  Prog.\ Theor.\ Phys.\  {\bf 106}, 873 (2001)
  and references therein.

\bibitem{shifman:anomaly}
  M.~A.~Shifman,
  Phys.\ Rept.\  {\bf 209}, 341 (1991)
  [Sov.\ Phys.\ Usp.\  {\bf 32}, 289 (1989\ UFNAA,157,561-598.1989)].

\bibitem{anomalymatching}
H.~Itoyama and A.~H.~Mueller,
  Nucl.\ Phys.\  B {\bf 218}, 349 (1983),
  R.~D.~Pisarski,
  Phys.\ Rev.\ Lett.\  {\bf 76}, 3084 (1996),
  R.~D.~Pisarski, T.~L.~Trueman and M.~H.~G.~Tytgat,
  Phys.\ Rev.\  D {\bf 56}, 7077 (1997),
S.~D.~H.~Hsu, F.~Sannino and M.~Schwetz,
  Mod.\ Phys.\ Lett.\  A {\bf 16}, 1871 (2001).

\bibitem{stern}
  M.~Knecht and J.~Stern,
  arXiv:hep-ph/9411253,
J.~Stern,
  arXiv:hep-ph/9712438,
  arXiv:hep-ph/9801282.

\bibitem{SDE}
B.~Holdom and G.~Triantaphyllou,
  Phys.\ Rev.\  D {\bf 51}, 7124 (1995);
  Phys.\ Rev.\  D {\bf 53}, 967 (1996),
B.~Holdom,
  Phys.\ Rev.\  D {\bf 54}, 1068 (1996).

\bibitem{SDE2}
P.~Maris and Q.~Wang,
  Phys.\ Rev.\  D {\bf 53}, 4650 (1996),
F.~S.~Roux, T.~Torma and B.~Holdom,
  Phys.\ Rev.\  D {\bf 61}, 056009 (2000).

\bibitem{shifman}
  I.~I.~Kogan, A.~Kovner and M.~A.~Shifman,
  Phys.\ Rev.\  D {\bf 59}, 016001 (1999).

\bibitem{watanabe}
  Y.~Watanabe, K.~Fukushima and T.~Hatsuda,
  Prog.\ Theor.\ Phys.\  {\bf 111}, 967 (2004).

\bibitem{skyrmion}
  M.~Rho,
  arXiv:0711.3895 [nucl-th];
  B.~Y.~Park, J.~I.~Kim and M.~Rho,
  Phys.\ Rev.\  C {\bf 81}, 035203 (2010).

\bibitem{hst}
  M.~Harada, C.~Sasaki and S.~Takemoto,
  Phys.\ Rev.\  D {\bf 81}, 016009 (2010),
  C.~Sasaki,
  arXiv:0910.4375 [hep-ph].

\bibitem{frankfurt}
  S.~Gallas, F.~Giacosa and D.~H.~Rischke,
  arXiv:0907.5084 [hep-ph].

\bibitem{hss}
  K.~Hashimoto, T.~Sakai and S.~Sugimoto,
  Prog.\ Theor.\ Phys.\  {\bf 120}, 1093 (2008).

\bibitem{takahashi}
  T.~T.~Takahashi and T.~Kunihiro,
  Phys.\ Rev.\  D {\bf 78}, 011503 (2008).

\end{thebibliography}
\end{document}